\documentclass[twocolumn, pre]{revtex4-1}

\usepackage{amssymb}
\usepackage{amsmath}
\usepackage{graphics}
\usepackage{graphicx}
\usepackage{verbatim}
\usepackage{mathcomp}
\usepackage{hyperref}

\newcommand{\be}{\begin{equation}}
\newcommand{\ee}{\end{equation}}
\newcommand{\WB}[1]{\overline{#1}}              

\begin{document}

\title{Scoring Strategies for the Underdog: A general, quantitative method for determining optimal sports strategies}

\date{\today}
\author{Brian Skinner}
\affiliation{Fine Theoretical Physics Institute, University of Minnesota, Minneapolis, Minnesota 55455}

\begin{abstract}

When facing a heavily-favored opponent, an underdog must be willing to assume greater-than-average risk.  In statistical language, one would say that an underdog must be willing to adopt a strategy whose outcome has a larger-than-average variance.  The difficult question is how much risk a team should be willing to accept.  This is equivalent to asking how much the team should be willing to sacrifice from its mean score in order to increase the score's variance.  In this paper a general, analytical method is developed for addressing this question quantitatively.  Under the assumption that every play in a game is statistically independent, both the mean and the variance of a team's offensive output can be described using the binomial distribution.  This description allows for direct calculations of the winning probability when a particular strategy is employed, and therefore allows one to calculate optimal offensive strategies.  This paper develops this method for calculating optimal strategies exactly and then presents a simple heuristic for determining whether a given strategy should be adopted.  A number of interesting and counterintuitive examples are then explored, including the merits of stalling for time, the run/pass/Hail Mary choice in American football, and the correct use of Hack-a-Shaq.

\end{abstract} \maketitle

\section{Introduction: Risk as a Statistical Quantity}

When winning is unlikely, a team must be willing to pursue risky strategies.  This is an important and generally well-accepted tenet of sports strategy \cite{Gladwell}.  Usually, this tenet manifests itself in the tendency of teams to attempt plays that have a low probability of success and a high potential yield in situations where the team faces a large deficit.  Conversely, heavily favored teams have a tendency to play conservatively, attempting to eliminate the possibility of an unlikely comeback by their opponent even at the cost of lowering their own final score.
 
In evaluating whether to use a given strategy, a coach/player must weigh the risk against the potential reward and decide whether adopting the strategy increases the team's chance of victory.  Specifically, the coach/player faces the following questions: How much risk should my team be willing to take?  By how many points should my team be trailing before it resorts to using a given high-risk/high-reward play?

It is the purpose of this paper to demonstrate that these types of strategic questions can be answered quantitatively by equating the notion of ``risk" with the statistical concept of variance in the team's final score.  A risky strategy, by definition, is one where there is a wider distribution of potential outcomes.  In statistical language, then, the idea of taking risk is that a team whose score has a lower expectation value than that of its opponent (the underdog) should be willing to pursue strategies that increase the variance in the final outcome, even at the cost of further lowering its expected final score.  In other words, sometimes the best strategy is the one that leads, on average, to a worse loss.

\begin{figure}[htb!]
\centering
\includegraphics[width=0.5\textwidth]{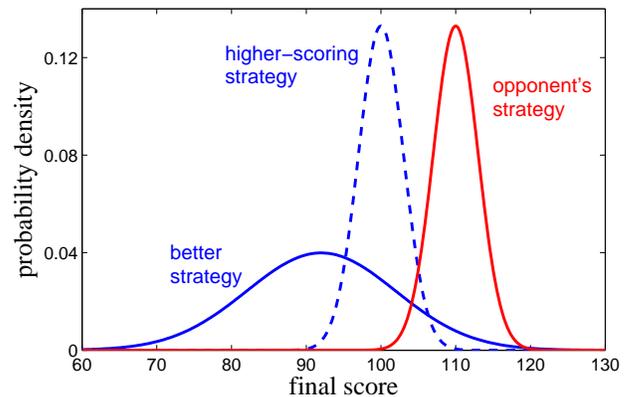}
\caption{(Color online)  Schematic portrayal of the distribution of final scores for two competing teams.  The underdog team (blue) improves their chance of winning if they pursue the strategy corresponding to the solid line rather than the dashed line, even though this lowers their expected final score.}
\label{fig:schematic}
\end{figure}

This idea is shown graphically in Fig.\ \ref{fig:schematic}, which shows the distribution in final score for two hypothetical teams.  A victory for the underdog team (blue curves) requires two independent occurrences: the underdog team must play unusually well, and their opponent must play unusually poorly.  The probability that the underdog will win is therefore represented graphically by the overlap between the tails of the distributions in the two teams' final scores.  In Fig.\ \ref{fig:schematic}, the solid blue line represents a better strategy for the underdog than the dashed blue line because this line has a greater overlap with the opponent's distribution (red curve).    

In quantitative language, the fundamental risk/reward tradeoff for an underdog team is between increasing the mean and increasing the variance of the team's final score.  In this rough sense, decisions that increase the overlap with the opponent's distribution are good ones, and decisions that decrease the overlap are bad \cite{Winningfootnote}.  The remainder of this paper is dedicated to showing how this overlap can be calculated for a given strategy using the binomial distribution.  A number of examples are developed to illustrate how the effectiveness of different ``risky" strategies can be evaluated both by an exact calculation and by a simple heuristic.

\section{Describing Strategies with the Binomial Distribution}

Given the simple reasoning of the previous section, it is tempting to discuss the value of ``streaky scorers" -- players whose offensive output varies significantly from game to game -- for overmatched teams.  Such players with great but inconsistent talent are a frequent source of discussion among fans, and, indeed, the insertion or removal of players who tend to ``get hot" would hypothetically be a method by which a team could control the variance of its final score.

Unfortunately, despite fans' anecdotal evidence, there is little statistical evidence for the existence of ``streaky" players.  Careful analysis in basketball, for example, has shown that the shooting patterns of essentially every NBA player can be described as a sequence of statistically independent shots \cite{Huizinga}.  If this is true -- that all scoring events in a game can be considered statistically independent -- then a team's final score is described by the binomial distribution.  More generally, when different scoring events have different point values, the final score is described by a product of binomial distributions, one for each type of scoring event.  In this case there is a strict relationship between the team's scoring percentages $\{p_i\}$, where $p_i$ is the probability of success of a given play $i$, and the variance $\sigma^2$ in its final score.  Specifically, running a play with success rate $p$ produces a variance
\begin{eqnarray}
\sigma^2 & = & (\textrm{point value of play})^2 \times  \label{eq:s2simple} \\
& & (\textrm{number of times play is run}) \times p \times (1-p) \nonumber.
\end{eqnarray}
Or, in more compact notation, $\sigma^2 = v^2 N p (1-p)$.  Here, $v$ denotes the point value of the play.  Since different plays $i$ contribute additively to the final score, the total variance in the final score is also additive:
\be 
\sigma^2 = \sum_i v_i^2 N_i p_i (1-p_i).
\label{eq:s2tot}
\ee

Eqs.\ (\ref{eq:s2simple}) and (\ref{eq:s2tot}) imply that a team seeking to alter the variance of its offensive strategy can take one of two approaches: they can try to run more plays with high point value $v$, or they can try to change the number $N$ of plays in the game \cite{intentionalmissfootnote}.  
Both of these approaches are considered through examples developed in Sec.\ \ref{sec:examples}.

For a given strategy, the increase in variance $\sigma^2$ should be weighed against the effect this strategy has on the team's mean score $\mu$, given by
\be 
\mu = \sum_i v_i N_i p_i.
\label{eq:mu}
\ee 
Generally speaking, the winning percentage corresponding to a given strategy can be calculated by integrating over the distributions corresponding to the team's final score and its opponent's final score.  This procedure is described in the Appendix, and is straightforward even though exact analytical expressions are cumbersome.  However, when dealing with strategic decisions that are not very short-term in nature ($N_i > 5$ or so for all $i$), optimum strategies can be calculated reliably from a simple heuristic based on the Central Limit Theorem (CLT).

\section{The CLT Rule} \label{sec:CLT}

In end-game situations, where only a very small number of plays remain in the game, it is fairly easy to calculate the probability of victory associated with a particular strategy by considering all possible outcomes of each play.  This summation procedure is formalized in the Appendix.  When the number of plays left in the game is large, however, it can be difficult to consider the net effect of every possible outcome.  Fortunately, in this limit one can invoke the CLT, which guarantees that for large $N$ the final differential score $\Delta = (\textrm{team's score}) - (\textrm{opponent's score})$ is Gaussian-distributed with mean $\mu - \mu_\textrm{opp}$ and variance $\sigma^2 + \sigma^2_\textrm{opp}$, where the subscript ``opp" labels the mean and variance of the opponent.  The probability $P$ of winning is therefore equal to the probability that $\Delta > 0$, which is given by
\be 
P \simeq \frac{1}{2} \left[ 1 + \textrm{erf}\left( \frac{Z}{\sqrt{2}} \right) \right] ,
\label{eq:Perf}
\ee 
where $\textrm{erf}(x)$ is the Gaussian error function and $Z$ is given by
\be 
Z = \frac{\mu - \mu_{opp}}{\sqrt{\sigma^2 + \sigma^2_{opp} } }.
\label{eq:CLTrule}
\ee

Since the win probability $P$ increases monotonically with $Z$, one can use Eqs.\ (\ref{eq:Perf}) and (\ref{eq:CLTrule}) to formulate the following simple rule, which for the remainder of this paper will be called the ``CLT Rule":

\vspace{-2mm}
\begin{center}
 \emph{A team's optimum long-term strategy is that which maximizes Z.}
\end{center}
\vspace{-2mm}

\noindent
Coupled with the descriptions of Eqs.\ (\ref{eq:s2tot}) and (\ref{eq:mu}) for calculating mean and variance, this rule gives a simple and powerful method for calculating optimal strategies.  Specifically, the CLT Rule suggests that if a team is trying to choose between pursuing strategy A and pursuing strategy B, then it can make that decision by evaluating the parameter $Z$, as defined in Eq.\ (\ref{eq:CLTrule}), for both strategies and then choosing the strategy that gives the higher value of $Z$.  The same procedure is also valid if the team is choosing between any larger number of distinct strategies: the strategy that gives the highest value of $Z$ is the best one.  Or, if the team is trying to optimize some continuous variable (for example, their average time of possession), then one can write $Z$ as a function of whatever strategic variable needs to be optimized and then search for the maximum of the function \cite{findminfootnote}.

The following section is dedicated to developing a number of examples from basketball and American football that illustrate the nature of the mean/variance tradeoff in sports and the effectiveness of the CLT Rule.  For clarity's sake, these examples are somewhat simplified, but there is in principle no reason why one cannot use the same procedure they describe to construct more detailed examples that take into account any number of additional factors, such as turnover rates, probability of fouls or penalties, or an increased number of offensive options.  Sec.\ \ref{sec:conclusion} briefly discusses potential applications to other sports, including tennis, golf, baseball, and soccer.

\section{Examples} \label{sec:examples}

\subsection{Shooting 3's in basketball} \label{sec:shoot3s}

In basketball, the most straightforward method of adopting risk is through the three-point shot.  As an example, consider a team that shoots two-pointers with a percentage $p_2 = 0.5$ and three-pointers with a percentage $p_3 = 0.\WB{33}$.  Such a team will have the same average score regardless of whether it chooses to shoot 2's or 3's, but in the latter case its variance will be twice as large [by Eq.\ (\ref{eq:s2simple})].  Using the logic from previous sections, we can immediately conclude that the team should shoot 3's when its expected final score is lower than that of its opponent, and it should shoot 2's when it is favored to win.

In more realistic examples, a team that resorts to shooting only 3's is likely to see its mean score decrease \cite{3sfootnote}.  
Suppose, for example that a team shoots 2's and 3's with percentages $p_2 = 0.5$ and $p_3 = 0.3$, respectively, and is facing a better opponent that shoots only \cite{shoot3sfootnote}
 2's with rate $p_\textrm{opp} = 0.55$.  In this case, by Eqs.\ (\ref{eq:s2tot}) and (\ref{eq:mu}), the team sacrifices 10\% from its expected score by switching to the three-point shot but increases its variance by 90\%.  In which situations is this tradeoff favorable?

\begin{figure}
\centering
\includegraphics[width=0.5\textwidth]{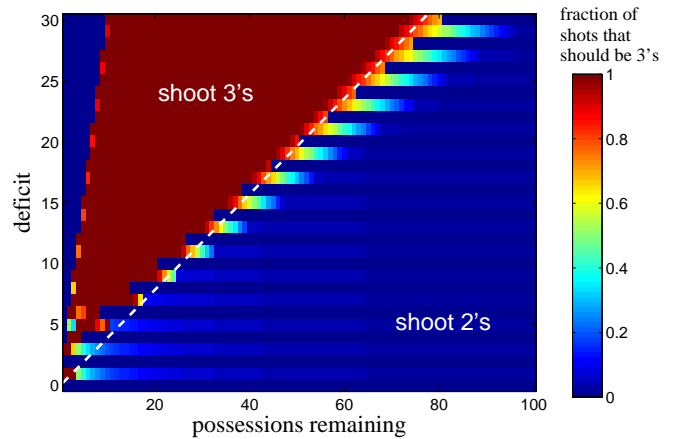}
\caption{Color online)  The optimal rate of 3-point shooting for a team with $p_2 = 0.5$ and $p_3 = 0.3$ against an opponent with $p_2 = 0.55$.  The dashed white line shows the prediction of the CLT Rule for when the team should shoot 3's instead of 2's.  The top left corner of this plot (dark blue) corresponds to situations where the team is mathematically eliminated.}
\label{fig:shoot3s}
\end{figure}

The answer to this question can be calculated exactly using the binomial distribution, as described in the Appendix.  This procedure allows one to determine the optimal rate of 3-point shooting as a function of the number of possessions remaining in the game, $N$, and the deficit $s$ that the team is facing.  The result is plotted in Fig.\ \ref{fig:shoot3s}.  

The main features of Fig.\ \ref{fig:shoot3s} can easily be understood by using the CLT Rule.  The simplest approach is to evaluate the value of the parameter $Z$ associated with 3-point shooting, $Z_3(N,s)$, as well as the one associated with 2-point shooting, $Z_2(N,s)$.  At values of $N$ and $s$ for which $Z_3$ is larger than $Z_2$, the team should shoot 3's.  If one equates $Z_3(N,s)$ and $Z_2(N,s)$, one arrives at the Eq.\ $s_{2/3} = 0.39 N$, which describes the optimal transition point from 2- to 3-point shooting.  This result is plotted as the dashed line in Fig.\ \ref{fig:shoot3s}.  Roughly speaking, when the deficit $s > s_{2/3}$ at a given $N$ the team should shoot 3's.  This prediction of the CLT rule is in excellent agreement with the exact calculation when $N > 10$ or so.

\subsection{The Run, the Pass, and the ``Hail Mary" in Football} \label{sec:rpHM}

\begin{figure*}
\centering
\includegraphics[width=0.85\textwidth]{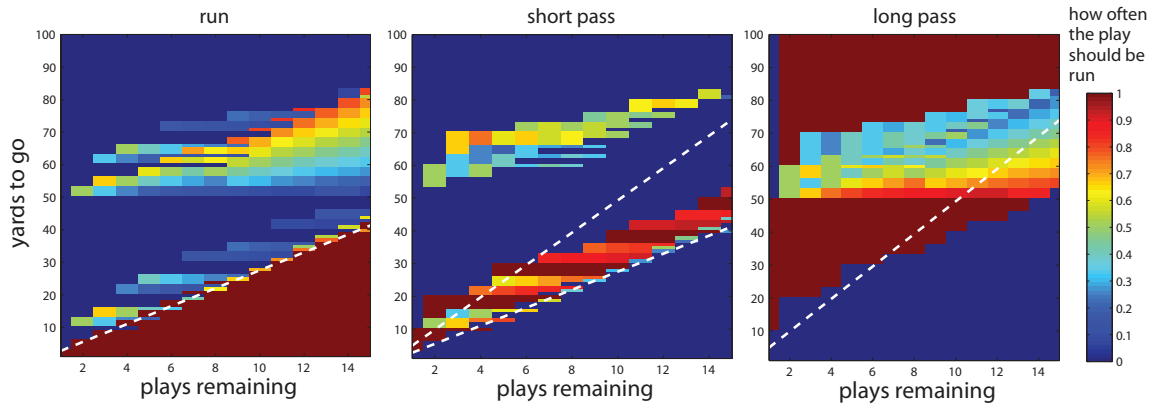}
\caption{Color online)  The optimal usage rate of each play for a football team that can attempt a 3-yard run, a 10-yard pass, and a 50-yard pass, with success rates 0.9, 0.25, and 0.02, respectively.  Dashed white lines show predictions from the CLT Rule.}
\label{fig:3plays}
\end{figure*}

In American football, a team has a much wider range of risk-taking possibilities on offense, since different plays can have a very different expected yield in terms of yards gained.  The question of ``which play should the team run?" is therefore a more complicated one.  As a simple example, however, we can consider the problem of a team that needs to move $y$ yards in the next $N$ plays \cite{4downfootnote}.
Suppose that the team is choosing between three play options: a short run that will yield 3 yards, a short pass that will yield 10 yards, and a long ``Hail Mary" pass that will yield 50 yards.  For the sake of example, assume that these plays have success rates 0.9, 0.25, and 0.02, respectively, so that the run produces the most yards on average while the passing plays offer different levels of risk/reward.  

In this case one can evaluate exactly which combination of play calls will yield the highest probability of winning.  The result is plotted in Fig.\ \ref{fig:3plays}.  The main features of this plot can be understood by performing the same simple analysis as described in Sec.\ \ref{sec:shoot3s}.  That is, one can evaluate the parameters $Z_\textrm{run}$, $Z_\textrm{short}$, and $Z_\textrm{long}$ associated with using the run, short pass, and long pass, respectively, assuming that there are $N$ plays remaining in the game and the team needs to advance $y$ yards.  Equating $Z_\textrm{run}$ and $Z_\textrm{short}$ suggests that that team should run the ball whenever the number of yards to go is small enough that $y < 2.8 N$.  Similarly, equating $Z_\textrm{short}$ and $Z_\textrm{long}$ suggests that the team should throw the Hail Mary whenever $y > 4.9 N$.  In the intermediate range of $y$, such that $2.8 N < y < 4.9 N$, the short pass is generally the team's best option.  The two transition lines $y = 2.8 N$ and $y = 4.9 N$ are plotted as dashed white lines in Fig.\ \ref{fig:3plays}.

It should be noted that while the CLT Rule does an excellent job of predicting when a team should throw short passes instead of running the ball, its description of the transition between the short pass and the Hail Mary is less accurate.  This inaccuracy is a result of the very discrete, one-time nature of the Hail Mary play call: a team relying on the 50-yard pass generally needs just one of these passes to work, and it is not correct to use a large $N$ approximation for its success.  The regions of non-zero usage for run and short pass plays at $y > 50$ generally correspond to mixed Hail Mary/short play strategies, wherein the team hopes for a large gain with a single Hail Mary pass and then plays relatively conservatively once the big gain has been made.

\subsection{Stalling in Basketball} \label{sec:stalling}

It is in the interest of an underdog team to keep the game short.  For example, the Minnesota Timberwolves have almost no chance of beating the Los Angeles Lakers in a seven game series, but their chance of outscoring the Lakers in a given quarter is decent, and their chance of outscoring the Lakers in a given 60-second interval is almost 50\%.  One can therefore discuss strategies for the underdog which involve intentionally limiting the number of possessions in the game -- in other words, stalling for time.

Consider, for example, a team whose normal possession uses up 16 seconds of the shot clock and is successful 50\% of the time facing an opponent whose normal possession also takes 16 seconds but is successful 55\% of the time (for simplicity, in this example all shots are assumed to be 2's).  This team is considering whether to slow down their possessions intentionally, using the full 24 seconds with each possession and thereby reducing the total number of possessions in the game.  Under what circumstances should they do so?

\begin{figure}[b]
\centering
\includegraphics[width=0.45\textwidth]{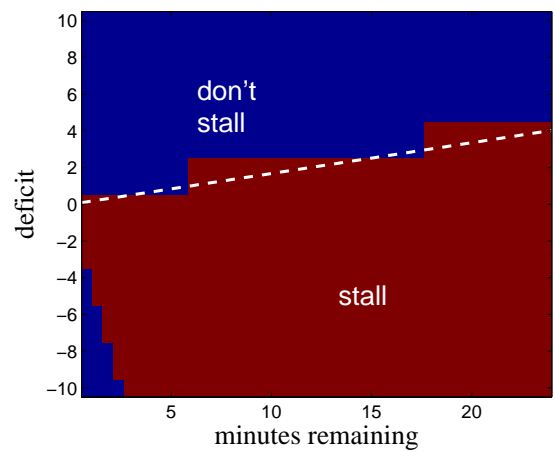}
\caption{Color online)  A diagram of when the team discussed in Sec.\ 4.3 should use ``stalling" as a tactic to increase their chance of beating a favored opponent.  Negative deficit implies a lead.  The dashed white line is the prediction of the CLT Rule.}
\label{fig:stall}
\end{figure}

This question can be answered, as in previous examples, by exact calculation and by the CLT Rule.  The result is shown in Fig.\ \ref{fig:stall}.  Its general conclusion -- that the team should stall when they have the lead and refrain from stalling when they don't -- is fairly intuitive.  Note, however, that the team may benefit from stalling \emph{even when they do not have the lead}, provided that the deficit they face is small and there is a significant amount of time remaining.   

Of course, this example assumes that the team's field goal percentage does not decline as a result of their stalling tactic.  If it does, then the slope of the CLT Rule line will decrease (and possibly become negative), so that the number of situations where stalling is favorable will diminish.

\subsection{Playing ``Hack-a-Shaq"} \label{sec:hackashaq}

Unlike the underdog, a favored team has an incentive to \emph{decrease} the variance in the game's final outcome.  One straightforward method of doing so is to ensure that the opponent takes only low-value shots.  This incentive can lead to ``Hack-a-Shaq" strategies in basketball, wherein the team fouls the opponent's worst free throw shooter in order to force him to shoot free throws rather than give the opponent an opportunity at a higher-value play \cite{HackaShaqfootnote}.

In fact, the favored team can benefit from ``Hack-a-Shaq" even when the fouled player's free throw percentage is not very low.  Consider, for example, that the variance of shooting $2N$ free throws with success rate $p$ is twice smaller than the variance of shooting $N$ two-pointers at the same rate.  A favored team can therefore benefit from fouling its opponent even when this does not lower the opponent's expected number of points scored.

\begin{figure}[h]
\centering
\includegraphics[width=0.45\textwidth]{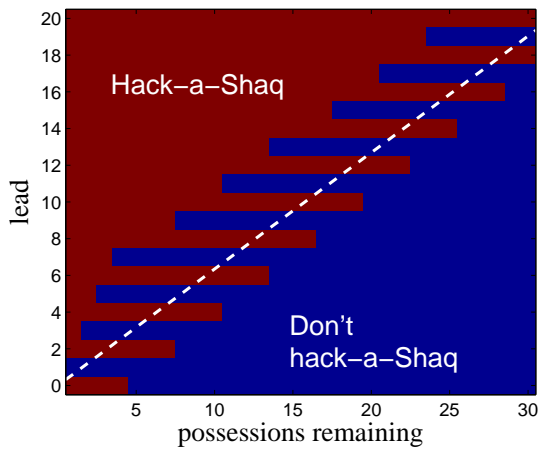}
\caption{Color online)  A diagram of when the favored team discussed in Sec. 4.4 should use ``Hack-a-Shaq" against their opponent.  The dashed white line is the prediction of the CLT Rule.}
\label{fig:hackashaq}
\end{figure}

Consider, for example, a team with (2-point) shooting percentage 55\% against a team with shooting percentage 50\% and free throw percentage 55\%.  Fig.\ \ref{fig:hackashaq} shows a diagram of the team's optimal use of the ``Hack-a-Shaq" strategy as a function of their lead and the number of possessions remaining \cite{combfootnote}.  
In this case, fouling increases the opponent's mean score (and over the long term, makes it equal to that of the fouling team).  Nonetheless, ``Hack-a-Shaq" is successful in increasing the favored team's chance of winning when it is done with a not-too-small lead and with a not-too-large amount of time remaining in the game.  

The idea that a team should foul intentionally when it has the lead remains somewhat controversial in basketball.  Its general rationale, however, can be stated simply using the following argument: If a team with the lead is given the choice of either giving its opponent 1 point or giving them a 50\% chance at 2 points, then it should give them the 1 point.  The mean score is the same either way, but the correct choice reduces the probability of an unlikely comeback.  In this argument, fouling intentionally is like giving up the 1 point: it reduces risk at a time when risk is the team's enemy.

\section{Optimization with Skill Curves} \label{sec:skillcurves}

In this section we return briefly to the examples of Sec.\ \ref{sec:shoot3s} and \ref{sec:rpHM} and comment on an important underlying assumption.  These examples, while illustrative, display in some places a fairly unrealistic result.  Namely, that under certain conditions the team should run exclusively a single type of play ({\it e.g.} shooting only 3's or using only running plays).  This extreme solution is a byproduct of the assumption that the success rate of a given play is independent of how often the play is run.  In fact, generally speaking, the more frequently a play is used, the less effective it will become as a result of increased defensive focus \cite{Oliver}.  This dependence amounts to a dependence of the success rate $p_i$ on the number of times $N_i$ that the play is run; such a dependency $p_i(N_i)$ is usually called a ``skill curve".  In cases where many plays remain in the game, skill curve relationships must be taken into account.  Their inclusion tends to eliminate the sharp boundaries between regions where different types of plays should be run and encourages mixed offensive strategies.

If the mathematical form of the skill curve relationship $p_i(N_i)$ is known, then the team's optimal strategy can be calculated using the exact method of the Appendix or by the approximate method presented in Sec.\ \ref{sec:CLT}.  One simply needs to calculate $p_i(N_i)$ for each offensive option $i$ before evaluating the win probability associated with the particular strategy $(N_1, ..., N_M)$.  Here each number $N_i$ in the list $(N_1, ..., N_M)$ labels the number of times that each of the $M$ plays is used.

When the total number of remaining plays $N = N_1 + ... + N_M$ in the game is very large, a team's best strategy is the one that optimizes their expected offensive efficiency $\mu$, and it is not necessary to consider the strategy's variance [note that, by Eqs.\ (\ref{eq:s2tot}) and (\ref{eq:mu}), the difference in expected efficiency is proportional to $N$ while the variance is proportional to $\sqrt{N}$, so that in the large $N$ limit optimizing $\mu - \mu_\textrm{opp}$ also optimizes $Z$].  A previous paper \cite{Skinner} described analytically how the dependence $p_i(N_i)$ can be incorporated into calculations of optimally-efficient offensive strategies.  Under the simplest approximation, the dependence $p_i(N_i)$ is a linear one:
\be 
p_i(N_i) = \alpha_i - \beta_i N_i/N,
\ee 
where $\alpha_i$ and $\beta_i$ are some numerical constants.  This form allows one to use the previous analytical results with a slight generalization for different play values $v_i$:
\be 
N_{i,\textrm{opt}} = N \frac{v_i \alpha_i + \lambda}{2 v_i \beta_i}.
\label{eq:Noptskillcurve}
\ee
Here, $\lambda$ is a constant (a Lagrange multiplier) given by
\be 
\lambda = \frac{2 - \sum_i (\alpha_i/\beta_i)}{\sum_i (v_i \beta_i)^{-1}}.
\ee

To illustrate the significance of this result and the effect of incorporating skill curve relations into calculations of optimum efficiency, one can consider an example very similar to that of Sec.\ \ref{sec:shoot3s} (and to the ``Ray Allen" example of Ref.\ \onlinecite{Skinner}, Sec.\ 3.4).  In particular, one can imagine a team that shoots with a constant percentage a $p_2 = 0.5$ on its two-pointers and a varying percentage $p_3(N_3) = 0.5 - 0.4 N_3/N$ on its three-pointers, where $N_3$ is the number of three-pointers taken by the team.  In this case, Eq.\ (\ref{eq:Noptskillcurve}) suggests that the optimal long-term strategy for the team is to have $N_3/N \approx 0.208$.  When there are few possessions remaining and the team faces a large deficit, however, $N_3$ should be larger in order to increase the variance of the team's final score.

\begin{figure}
\centering
\includegraphics[width=0.5\textwidth]{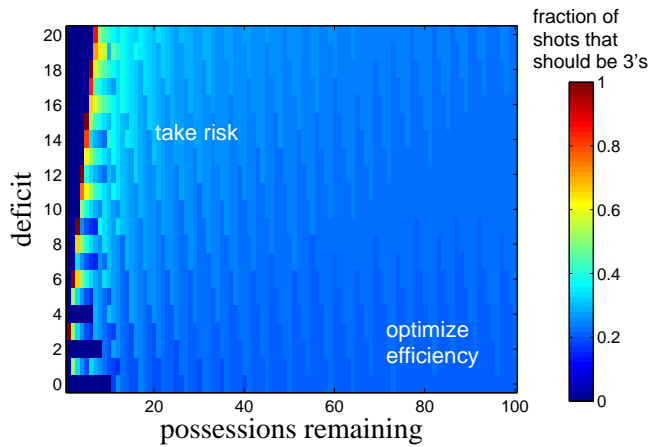}
\caption{Color online)  The optimal rate of 3-point shooting for a team with $p_2 = 0.5$ and $p_3 = 0.5 - 0.4 N_3/N$.  On the right side of the plot the solution closely approaches the result of Eq.\ (\ref{eq:Noptskillcurve}).}
\label{fig:shoot3sRA}
\end{figure}

Fig.\ \ref{fig:shoot3sRA} shows the result of an exact calculation using the method of the Appendix.  Note that in the large $N$ limit the result approaches that of Eq.\ (\ref{eq:Noptskillcurve}), while in regions with few possessions remaining and a large deficit the fraction of three-pointers increases markedly.  An approach based on the CLT Rule, where $Z$ is optimized as a function of the variable $N_3$, produces a very similar result as in Fig.\ \ref{fig:shoot3sRA}, although it is incapable of reproducing the quickly-varying behavior at very small $N$.

\section{Conclusion} \label{sec:conclusion}

This paper has presented a formal, quantitative method for optimizing risk/ reward decisions in sports and has applied it to a number of examples where ``taking risk" involves running low-probability/high-yield plays, stalling for time, or intentionally fouling.  
While the examples presented here have been drawn from basketball and American football, a number of applications to other sports can easily be imagined.  For example, in tennis the server is often much more conservative on his/her second serve than on the first, since the second serve carries an immediate risk of a double fault.  When the server is an underdog in the match, however, he/she should consider whether the potential reward of a more aggressive second serve is worth the risk it presents.  Similarly, in baseball a pitcher must decide how often to use risky pitches against a particular batter, and in this decision the game situation is an important factor.  In soccer and in golf players face similarly difficult risk/reward decisions, wherein they must decide whether to aggressively shoot for the goal/hole or whether to be more conservative and simply work to set up a better shot in the future.  In each of these sports, the decision of which strategy to pursue can be made quantitatively.

This paper's one crucial underlying assumption is that all plays in a game can be considered statistically independent of each other, so that the game's final score can be described using the binomial distribution.  While this assumption doesn't always sit well with fans and players, all advanced statistical studies thus far indicate that it is usually sufficiently close to the truth to be useful.  If its correctness can indeed be assumed, then the analyst is given a tremendous amount of predictive power based on the strict relationships between mean, variance, and play success rate that it implies [Eqs.\ (\ref{eq:s2simple}) - (\ref{eq:mu})].  Thus, in order to make quantitative predictions of the optimal strategy for a given game situation, it is necessary only to know the success rates $\{p_i\}$ of the offense's potential plays.  While measuring these rates is not necessarily an easy task, it is at least a straightforward one that leaves no room for subjective metrics and does not require the use of numeric simulations.

At the present moment there are many questions about describing sports performance for which current statistical methods are not yet sufficiently advanced that they can be trusted over the opinion of a trained observer of the game.  In these cases the role of sports statistics should be to point out things that such an observer has not noticed and to assist in making quantitative, strategic decisions.  It is in these latter situations -- where the coach or player thinks ``I know that I should be doing X, but I don't know to what extent I should do it" -- that analytical and theoretical approaches to sports statistics have a tremendous potential to be helpful.  The optimization method presented in this paper may provide a small step toward formalizing our sports intuition and using it to make quantitative decisions about optimum strategy.

\begin{acknowledgments}

The author would like to thank Henry Abbott, Ben Alamar, M. R. Goldman, Ryan Rodenberg, and the members of the APBRmetrics forum for helpful discussions.  This work is a revised version of a paper presented at the 2011 MIT Sloan Sports Analytics Conference.

\end{acknowledgments}

\appendix
\section{Exact Method for Calculating Optimum Strategies Using the Binomial Distribution} 

The binomial distribution $f(x; N)$ gives the probability that exactly $x$ out of $N$ plays will be successful, given that any one play has a chance $p$ of succeeding:
\be 
f(x;N) = \frac{N!}{x!(N-x)!} p^x (1-p)^{N-x}.
\label{eq:fbin}
\ee 
Consider a team that has some number $M$ of different plays they can run, and that a given play, labeled $i = 1, ..., M$, has a probability of success $p_i$ and produces $v_i$ points when it succeeds.  Then, if play $i$ is run a total number of $N_i$ times, the probability that each play $i$ will be successful exactly $x_i$ times is given by
\be 
f(x_1, ..., x_M; N_1, ..., n_M) = \prod_{i=1}^M f(x_i; N_i).
\ee 
The corresponding team score is
\be 
s(x_1, ..., x_M) = \sum_{i=1}^M v_i x_i .
\ee 
The distribution of the team's total score $s'$, denoted $F(s'; N_1, ..., N_M)$, that corresponds to a particular selection of play usage $(N_1, ..., N_M)$ can be found by summing over all possible outcomes $\{x_i\}$ for the team's plays:
\begin{widetext}
\begin{equation}
F(s'; N_1, ..., N_M) = \sum_{x_1 = 0}^{N_1} ... \sum_{x_M = 0}^{N_M} f(x_1, ..., x_M ; N_1, ..., N_M) \delta[ s(x_1, ..., x_M) - s'].
\label{eq:Fs}
\end{equation}
\end{widetext}
Here, $\delta[x]$ is the Kronecker delta function, defined by $\delta[0] = 1$ and $\delta[x \neq 0] = 0$.

If the opponent's strategy and use of plays is known, then the distribution $F(s_\textrm{opp})$ of their final score $s_\textrm{opp}$ can also be calculated as described by Eqs.\ (\ref{eq:fbin}) -- (\ref{eq:Fs}).  Then the chance $P$ that the team will win (or tie) when employing the strategy defined by the shot selection $(N_1, ..., N_M)$ is calculated using a sum over all possible scores of the team and its opponent:
\be 
P(N_1, ..., N_M) = \sum_s \sum_{s_\textrm{opp}} F(s; N_1, ..., N_M) F(s_\textrm{opp}) \Theta(s - s_\textrm{opp}).
\ee 
Here, $\Theta(x)$ is the Heaviside step function, defined by $\Theta(x <0) = 0$ and $\Theta(x \geq 0) = 1$.

The team's optimum strategy is the one whose set of shots $(N_1, ..., N_M)$ maximizes the win probability $P(N_1, ..., N_M)$.  In the results presented in Sec.\ \ref{sec:examples} of this paper, the optimum strategy is found by evaluating $P(N_1, ..., N_M)$ at every possible combination $(N_1, ..., N_M)$, subject to the constraint that the sum of $N_i$ is equal to the total number of remaining possessions.  While this calculation of an optimum strategy can involve many distinct summations, it is not particularly taxing computationally.  For $N < 100$ or so and $M < 5$, the calculation can generally be done in real time.

It should be noted, finally, that the binomial description of basketball game scores has been employed by previous authors \cite{Oliver2}, although usually in a somewhat simplified version that characterizes all possessions as ``successes" or ``failures" and does not allow for plays with different point value $v$.



\begin{thebibliography}{100}
\bibitem{Winningfootnote} Of course, it is possible that a sports coach might not simply equate ``good/bad" with ``higher/lower chance of winning".  For example, the strategy corresponding to the dashed curve in Fig.\ \ref{fig:schematic} gives the team a smaller chance of winning, but it also reduces the likelihood of a truly embarrassing loss that could cost the coach his job.  The psychological aspects associated with a blowout loss are ignored here, but they might in fact be essential to understanding real-world strategic decisions in sports.
\bibitem{findminfootnote} In many cases, this process can easily be done analytically by taking the derivative of $Z$ with respect to the strategic variable $x$ and then setting $dZ/dx$ equal to zero.  In more complicated cases, where multiple strategic variables are being optimized simultaneously, one can usually use straightforward gradient ascent algorithms to find the optimum.
\bibitem{3sfootnote} While this is likely true, it is also the true that every NBA team currently scores more with the average 3-point shot than with the average 2-point shot (basketballreference.com), a difference that cannot be explained by the slightly lower rates of offensive rebounding off of 3-point misses.  It is tempting, then, to conclude that all NBA teams should be shooting drastically more 3's.  However, one should recall that the effectiveness of a given play declines with its usage rate \cite{Oliver}, and so a basketball team has an incentive to use its 3-point shot sparingly and thereby maintain a difference between 2-point and 3-point scoring rates \cite{Skinner}.  See Sec.\ \ref{sec:skillcurves} for additional relevant discussion.
\bibitem{shoot3sfootnote} Since this opponent has a larger expected score, it is to their advantage to shoot only 2's in order to minimize variance.
\bibitem{4downfootnote}  The unique, four-down structure of football is ignored in this example.  For example, setting $N = 10$ and $y = 40$ is assumed to mean ``the team needs to advance a total of 40 yards in the next 10 plays", and not the somewhat different requirement ``the team needs to advance at least 10 yards during any four consecutive plays and at least 40 yards during the next 10 plays."
\bibitem{HackaShaqfootnote} This use of ``Hack-a-Shaq" should be distinguished from the one used by trailing teams in end-game situations, whose goal is to increase the number of possessions remaining and not to alter their opponent's mean score or variance.  Such situations can also be analyzed by the methods presented in this paper.
\bibitem{combfootnote} The unusual ``comb" structure of the border between the two regions in Fig.\ \ref{fig:hackashaq} is a consequence of the assumption that only 2-point shots are taken.  This places a significant difference between odd-numbered and even-numbered leads.  In more complex examples, where free throws and three-pointers are included, this structure disappears in the limit of many remaining possessions.
\end{thebibliography}
\end{document}